**Indentation of a Rigid Sphere into an Elastic Substrate with Surface Tension and Adhesion**

Chung-Yuen Hui[1*], Tianshu Liu[1], Thomas Salez[2], Elie Raphael[3], Anand Jagota[4]

[1]Field of Theoretical and Applied Mechanics, Department of Mechanical and Aerospace Engineering, Cornell University, Ithaca, NY 14850, [1*]corresponding author: ch45@cornell.edu

[2]School of Engineering and Applied Sciences, Harvard University, Cambridge, MA 02138

[3]PCT Lab, UMR CNRS 7083 Gulliver, ESPCI ParisTech, PSL Research University, 10 rue Vauquelin, 75005, Paris, France

[4]Department of Chemical and Biomolecular Engineering, Lehigh University, Bethlehem, PA

**Abstract**

The surface tension of compliant materials such as gels provides resistance to deformation in addition to and sometimes surpassing that due to elasticity. This article studies how surface tension changes the contact mechanics of a small hard sphere indenting a soft elastic substrate. Previous studies have examined the special case where the external load is zero, so contact is driven by adhesion alone. Here we tackle the much more complicated problem where, in addition to adhesion, deformation is driven by an indentation force. We present an exact solution based on small strain theory. The relation between indentation force (displacement) and contact radius is found to depend on a single dimensionless parameter: $\omega = \dfrac{\sigma}{(\mu R)^{2/3}\left(\dfrac{9\pi}{4}W_{ad}\right)^{1/3}}$ , where $\sigma$ and $\mu$ are the surface tension and shear modulus of the substrate, $R$ is the sphere radius, and $W_{ad}$ is the interfacial work of adhesion. Our theory reduces to the Johnson-Kendall-Roberts theory and Young-Dupre equation in the limits of small and large $\omega$ respectively and compares well with existing experimental data. Our results show that, although surface tension can significantly affect the indentation force, the magnitude of the pull-off load in the partial wetting liquid like limit is reduced only by 1/3 compared with the JKR limit and the pull-off behavior is completely determined by $\omega$.

## 1  Introduction

A fundamental problem in adhesion science is the contact between a rigid sphere and an elastic substrate. For small deformation where the contact radius $a$ is small in comparison with the radius of the sphere $R$, the solution was derived by Hertz [1]. Hertz theory assumes adhesion-less contact; as a result, the traction on the contact region is compressive everywhere and vanishes at the contact line. In 1971, Johnson, Kendall and Roberts (JKR) [2] extended Hertz theory to allow for adhesion. In essence, they treated the contact line as the tip of an external crack; the equilibrium relation between the applied load and the contact radius was obtained by equating the energy release rate of this crack to the interfacial work of adhesion.

The JKR theory has been extremely successful in describing the adhesive contact of elastic spheres. It is therefore surprising to find that recently observed deformations of soft substrates, such as plasticized polystyrene [3], hydrogels [4] and silicone gels [5] caused by adhesion of hard microparticles or nanoparticles, in the absence of external load, deviate considerably from JKR theory. For example, Style *et al.* [5] have reported that the power-law relation between the contact radius $a$ (indentation depth $\delta$) and sphere radius $R$ changes from $a \propto R^{2/3}$ $(\delta \propto R^{1/3})$ to $a \propto R$ $(\delta \propto R)$ as the sphere reduces in size or the substrate becomes softer. The transition in scaling observed in these experiments has been interpreted as a corresponding underlying transition from the JKR limit where the adhesion-driven deformation is primarily resisted by bulk elasticity, to the "liquid" limit where the adhesion-driven deformation is primarily resisted by the substrate-air surface tension.

Style *et al.* [5] noted that these deviations can be explained by the fact that JKR theory does not account for the role of the substrate-air surface tension in resisting deformation. More specifically, in JKR theory, the work done by the surface tensions upon change in surface area is neglected in the calculation of the energy release rate. Since decreasing the substrate elastic modulus or the sphere radius increases the relative contribution of surface tension to the energy release rate, it is not surprising that the JKR theory breaks down for sufficiently soft substrates or small spheres.

Using a large deformation Finite Element Model which incorporates both substrate–air surface tension and nonlinear-elasticity, Xu *et al.* [6] have computed the energy release rate for the problem of a rigid sphere in adhesive contact with a neo-Hookean substrate in the absence of external load. Their numerical results show a transition from the elasticity dominated regime where $a \propto R^{2/3}$ to the surface tension dominated regime where $a \propto R$. This transition depends on the single elasto-capillary number $\alpha \equiv \sigma / (2\mu R)$ - small $\alpha$ favors elasticity whereas large $\alpha$ favors surface tension. Their results are found to be in good agreement with the experiments reported by Style *et al.* [5]. This transition in scaling has been verified using molecular dynamics simulations by Cao *et al.* [7].

Although hard particles are used in the above reported experiments, a similar transition is expected for soft particles on a rigid surface. For example, Lau *et al.* reported some preliminary observation of such effects with spontaneous adhesion of latex nanoparticles [8]. The transition from the JKR limit to the liquid limit for a soft sphere on a rigid substrate in the absence of external load was also studied by Salez *et al.* [9] using an ad-hoc thermodynamical approach. These works illustrate that the departure from JKR scaling represents the increasing influence of solid surface stress in resisting the adhesion-driven deformation.

The spheres in the aforementioned experiments and theories are subjected to zero external load. This is a special case of the more general situation in which deformation and contact are driven by some combination of external load and work of adhesion at the contacting interface. The solution for Hertzian contact between a rigid cylinder and an elastic half-space with indentation force was given by Long *et al.* [10]. For the case of a soft sphere on a rigid half space, a molecular dynamics simulation and a scaling analysis of this problem were reported by Carrillo and Dobrynin [11]. The theoretical analysis is much more complicated if an external load is applied, and the goal of this article is to provide a rigorous analysis of this problem using a continuum mechanics approach.

2.  **Statement of Problem and Summary of Approach**

Our system is shown in Fig. 1. A rigid sphere of radius $R$ is brought into contact with the flat surface of a substrate by a vertical force $P$. The substrate initially occupies the half-space z > 0 and is assumed to be linearly elastic, isotropic and incompressible, with shear modulus $\mu$ and Poisson's ratio v = 0.5. Just as in JKR theory, our analysis is based on small deformation theory where the displacements and strains are small. We also assumed that the surface stress tensor is isotropic, so it can be represented by the scalar surface tension $\sigma$. Since the materials with which we are concerned are isotropic in the plane of the surface, the surface stress is expected to be isotropic as well and can be represented by a single number, $\sigma$, the scalar surface tension. This assumption has also been made by a number of other investigators and is validated by the good agreement between the resulting models and experiments [12]. Note that, in general, the surface tension of a solid does not need to be the same as its surface energy. For a detailed discussion, consult [13,14].

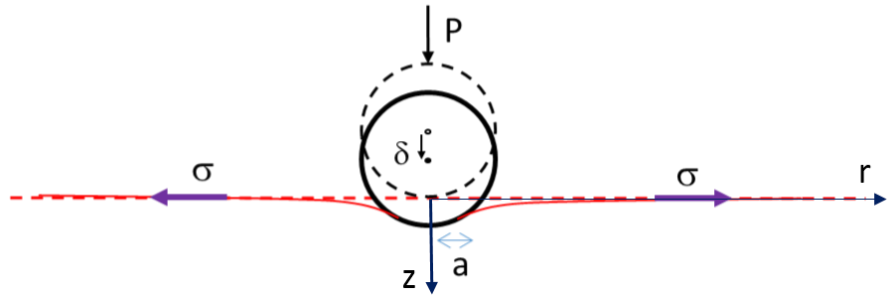

Fig. 1 A rigid sphere of radius $R$ is indented by a vertical force $P$, bringing it into no-slip contact with the surface of an incompressible isotropic elastic half-space. The indentation is resisted by the elasticity of the substrate as well as the substrate-air surface tension $\sigma$.

In contrast to JKR theory, which assumes frictionless contact, we assume a no-slip contact condition. That is, a material point on the substrate surface is held fixed once it comes into contact with the rigid sphere. Under this boundary condition, we only need to specify the surface tension of the non-contacting part of the substrate surface. Without the no-slip condition, one would have to introduce a new parameter into the model – the interfacial surface tension in the contact region. In general, the no-slip contact condition will induce shear stress in the contact region. However, a classical result in linear elasticity states that the no-slip boundary condition is consistent with a vanishing shear stress provided that the sphere is rigid and the substrate is incompressible and infinite in extent [1]. Fortunately, for most elastomers and hydrogels, incompressibility is an excellent approximation.

As in JKR theory, our approach is based on energy balance, that is, the relation between the contact radius and the applied load is obtained by setting the energy release rate $G$ of the external crack (the air gap between the sphere and the surface of the substrate outside the contact zone) equal to the interfacial work of adhesion, $W_{ad}$. The key novel feature is the inclusion of the substrate-air surface tension in the calculation of $G$.

Suppose we are able to obtain the "Hertz-like" solution of the contact problem, that is, the relation between force, $P_H(a)$, indentation displacement, $\delta_H(a)$, and contact radius, $a$, in the absence of any adhesive forces. Note that the subscript $H$ in $P_H$ and $\delta_H$ refers to adhesion-less contact, and these quantities in general depend on the air-substrate surface tension. For the special case of zero surface tension, $P_H$ and $\delta_H$ reduce to the classical Hertz solution. Having this solution automatically also means that we know the contact compliance,

$$C(a) = d\delta_H / dP_H = \frac{d\delta_H}{da} / \frac{dP_H}{da}. \tag{1}$$

This information suffices to determine the change of potential energy of the system (not counting interfacial work of adhesion) per unit change in contact area, i.e., the energy release rate, as a function of contact area. Specifically, as previously shown by Vajpayee et al. [15], the energy release rate, $G$, is given by

$$G = -\frac{1}{4\pi a}\frac{dC}{da}(P - P_H)^2 = -\frac{1}{4\pi aC^2}\frac{dC}{da}(\delta - \delta_H)^2. \tag{2a,b}$$

where $P$ and $\delta$ are respectively the applied load and indentation depth (see Fig. 1) of the adhesive contact problem.

## 2.1 The JKR limit

As an example, consider the classical Hertz theory without surface tension [1],

$$P_H = \frac{16\mu a^3}{3R}, \quad \delta_H = \frac{a^2}{R} \Rightarrow C = \frac{1}{8\mu a}. \tag{3a}$$

Substituting (3a) into (2a) and enforcing the energy balance equation $G = W_{ad}$ recovers the JKR theory for a rigid sphere in adhesive contact with an incompressible elastic half-space,

$$P = \frac{16\mu a^3}{3R} - 4\sqrt{2\pi\mu a^3 W_{ad}}. \tag{3b}$$

The indentation depth versus contact radius relation can be obtained using (2b) and (3a):

$$\delta = \frac{a^2}{R} - \sqrt{\frac{\pi a W_{ad}}{2\mu}}. \tag{3c}$$

Thus, the main thrust of our analysis is to determine how the Hertz-like load and Hertz-like displacement (and hence the instantaneous compliance) are affected by the presence of substrate-air surface tension.

## 2.2 The surface tension dominated or the partial wetting liquid-like limit.

Before tackling the general problem of the transition from the regime where resistance to deformation is dominated by elasticity (eqns. 3a-c) to the surface tension dominated limit, we examine the limiting "liquid-like" case in which substrate-air surface tension dominates over elasticity. The free-body diagram in Fig. 2 shows that force balance requires

$$P = 2\pi \sigma a \sin(\theta - \theta_p), \tag{4}$$

where $\theta = \sin^{-1}(a/R)$ and $\theta_p$ is the peel angle. The energy release rate $G$ can be computed using Kendall's peel theory [16]:

$$G = \sigma(1 - \cos\theta_p). \tag{5}$$

By setting $W_{ad} = \sigma + \gamma_{RA} - \gamma_{RS}$, the energy balance equation $G = W_{ad}$ is equivalent to the Young-Dupre equation

$$\gamma_{RS} + \sigma \cos(\pi - \theta_p) = \gamma_{RA}, \tag{6}$$

where the $\gamma$'s are the surface energies and where the subscripts $R, S, A$ stand for rigid sphere, compliant substrate and air atmosphere respectively. Note that in (6) we have used surface energies instead of surface tensions, since the Young-Dupre equation is based on energy balance; for a detailed discussion see [17]. Note also that in the liquid-like limit, we have assumed the surface tension and surface energy of the air-substrate interface to have the same value. Using the small angle approximation, the energy balance equation becomes

$$\sigma \theta_p^2 / 2 = W_{ad}. \tag{7}$$

Substituting (7) into (4) and using the small angle approximation $\sin\theta_p \approx \theta_p$ gives the relation between applied load and contact radius in the liquid limit:

$$P = \frac{2\pi\sigma a^2}{R} - 2\pi a\sqrt{2\sigma W_{ad}}. \tag{8}$$

Equation (8), which, when specialized to the zero force case, gives a contact radius of

$$a_0^{Liquid} = R\sqrt{2W_{ad}/\sigma}, \tag{9}$$

that is directly proportional to the sphere radius $R$. The scaling predicted by (9) is consistent with the experiments of Style *et al.* [5] who showed that when a small sphere is placed on the surface of a soft gel (with practically zero external force), the contact radius scales linearly with the radius instead of to the power 2/3 predicted by JKR theory.

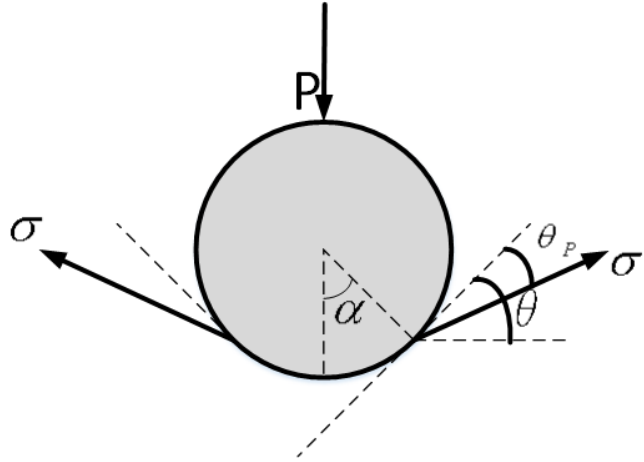

Fig. 2: Free body diagram showing the forces acting on the sphere in the partial wetting liquid-like limit.

### 2.3  Comparison between the JKR limit and the partial wetting liquid-like limit.

The effect of surface tension is most evident by comparing the JKR limit with the partial wetting liquid-like limit by plotting (3b) and (8) in Fig. 3(a,b). As noted by Chaudhury *et al.* [18], in the JKR limit a plot of $(a/R)^{3/2}$ versus $P/\left(4\sqrt{2\pi\mu W_{ad}a^3}\right)$ gives a straight line with slope $3/\left(4\sqrt{\mu R/(2\pi W_{ad})}\right)$ and horizontal intercept $-1$ (see insert of Fig. 3a). In the surface tension dominated limit, this plot is not a straight line. Instead, as shown in the insert of Fig. 3b, $a/R$ versus $P/(2\pi a\sigma)$ is a straight line with slope 1 and horizontal intercept $-\sqrt{2W_{ad}/\sigma}$. In a load controlled test [1], the sphere jumps out of contact at a contact radius of

$$a_{off}^{Elastic} = \frac{1}{2}\left(\frac{9\pi W_{ad}}{4\mu}\right)^{1/3} R^{2/3} \,. \tag{10a}$$

The pull-off load at this instability is

$$P_{off}^{Elastic} = -\frac{3}{2}\pi R W_{ad} \,. \tag{10b}$$

The corresponding quantities in the surface tension dominated limit are:

$$a_{off}^{Liquid} = \sqrt{\frac{W_{ad}}{2\sigma}} R \,, \tag{11a}$$

$$P_{off}^{Liquid} = -\pi W_{ad} R, \qquad (11b)$$

respectively. Comparing (10b) and (11b), one sees that in the liquid limit the magnitude of the pull-off load is reduced by 1/3 compared with the JKR limit. More importantly, (10a) and (11a) show that the contact radius at pull-off scales very differently with the sphere radius, despite the fact that the scaling for the pull-off forces is the same.

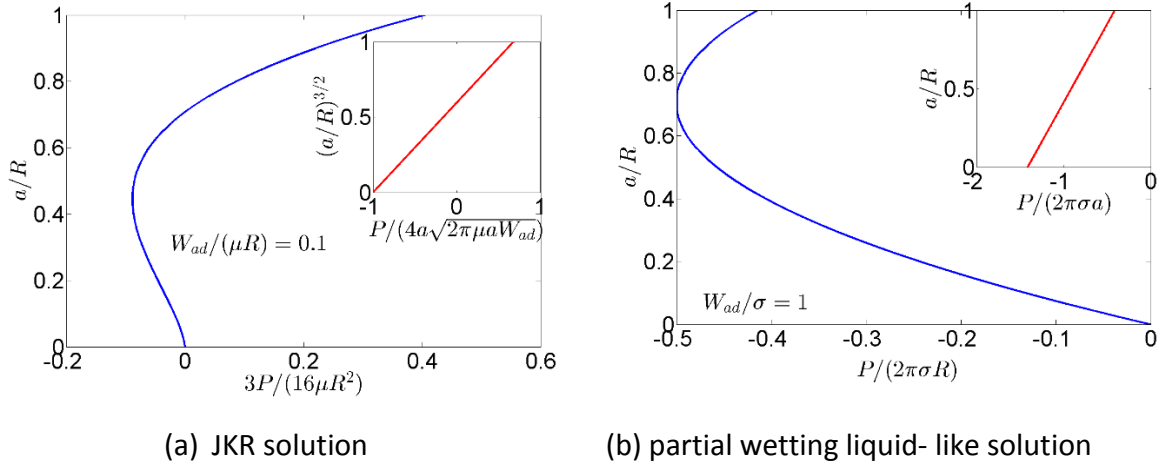

(a) JKR solution  (b) partial wetting liquid-like solution

Fig. 3(a), (b) Normalized contact radius versus normalized indentation load for the JKR limit and the partial wetting liquid-like limit. Insets show different scaling behaviors. Note that load is normalized differently in (a) and (b). The inserts in these figures show different scaling behaviors.

### 3. General case: transition between JKR and partial wetting liquid-like limits:

#### 3.1 Determination of $P_H, \delta_H$ and C

The adhesion-less contact of a rigid sphere to an infinite elastic substrate with surface tension has recently been studied by Long et al. [19]. Here we use their formulation specialized to an incompressible elastic substrate. For reasons stated in the 2nd paragraph of section 2.1, their formulation is strictly valid only for this case. Their numerical results are presented for specific values of sphere radius, applied load and material properties. Since our goal is to determine the relation between applied load and contact radius for all material and loading parameters, a different approach is needed.

Using the point source solution of Hajji [20], Long et al. [19] showed that the contact pressure $p$ acting on the sphere satisfies the integral equation:

$$\frac{1}{2\sigma}\int_0^a tp(t)\int_0^\pi \psi(l/s)d\theta dt = \delta_H - \frac{r^2}{2R} \qquad r < a \qquad (12a)$$

where

$$l = \sqrt{r^2 + t^2 - 2rt\cos\theta}, \quad s = \sigma/2\mu, \tag{12b}$$

$$\psi(\eta) = H_0(\eta) - Y_0(\eta). \tag{12c}$$

Here $(r,\theta)$ is the polar coordinate system on the surface of the substrate with origin at the initial contact point, $\delta_H$ is the displacement of the rigid sphere in the absence of adhesion, and $H_0, Y_0$ are the Struve function and Bessel function of the 2$^{nd}$ kind (both of order 0), respectively. By setting $r = 0$ in (12a), $\delta_H$ is found to be:

$$\delta_H = \frac{\pi}{2\sigma}\int_0^a tp(t)\psi(t/s)dt. \tag{13a}$$

Substituting (13a) into (12a) allows us to eliminate $\delta_H$ and the integral equation becomes:

$$\frac{\pi}{2\sigma}\int_0^a tp(t)\left[\psi(t/s) - \frac{1}{\pi}\int_0^\pi \psi(l/s)d\theta\right]dt = \frac{r^2}{2R} \quad r < a. \tag{13b}$$

Force balance implies that the load $P_H$ acting on the sphere is related to the pressure distribution by

$$2\pi \int_0^a tp(t)dt = P_H. \tag{13c}$$

To reduce the number of material parameters in the calculation, we introduce the following normalization:

$$r = a\bar{r}, t = a\bar{t}, l = a\bar{l}, p = \frac{8\mu a}{\pi R}\bar{p}, P_H = \frac{16\mu a^3}{3R}\bar{P}_H, \delta_H = \frac{a^2}{R}\bar{\delta}_H \tag{14}$$

Thus, all length scales are normalized by the unknown contact radius *a*. In particular, the total load (Hertz-like load) and the displacement of the sphere (Hertz-like displacement) are normalized by the Hertz load and displacement of the classical Hertz theory without surface tension, respectively (see (3a)). Thus, in the absence of surface tension, the *normalized* Hertz-like load $\bar{P}_H$ and displacement $\bar{\delta}_H$ are both exactly equal to one.

Our normalization yields a very important result. The integral equation (13b), the Hertz-like displacement (13a) and load (13c), after normalization, become:

$$2\beta^{-1}\int_0^1 \bar{t}\bar{p}(\bar{t},\beta)\left[\psi(\bar{t}/\beta) - \frac{1}{\pi}\int_0^\pi \psi(\bar{l}/\beta)d\theta\right]d\bar{t} = \frac{\bar{r}^2}{2} \quad \bar{r} < 1 \tag{15a}$$

$$\bar{P}_H(\beta) = 3\int_0^1 \bar{t}\bar{p}(\bar{t},\beta)d\bar{t} \tag{15b}$$

$$\bar{\delta}_H(\beta) = \frac{2}{\pi\beta}\int_0^1 \bar{t}\bar{p}(\bar{t},\beta)\psi(\bar{t}/\beta)d\bar{t} \tag{15c}$$

$$\beta = \sigma/(2\mu a). \tag{15d}$$

The important result is that the normalized Hertz-like load and displacement depend on a *single* dimensionless parameter $\beta$, which is the ratio of the elasto-capillary length $\sigma/\mu$ to the contact diameter. Physically, a small $\beta$ corresponds to the classical Hertz limit, where elasticity dominates, whereas a large $\beta$ corresponds to the non-wetting liquid limit, where surface tension dominates.

Although useful from a theoretical standpoint, the integral equation (15a) is not suitable for numerical solution. Numerical solution is carried out using the equivalent form

$$\frac{2}{\pi\beta}\int_0^1 \bar{t}\bar{p}(\bar{t},\beta)\int_0^\pi \psi(\bar{l}/\beta)d\theta d\bar{t} = \bar{\delta}_H - \frac{\bar{r}^2}{2} \quad \bar{r}<1. \tag{16}$$

Details of our numerical scheme are given in the supplementary material. Briefly, for a fixed value of $\beta$, we first set $\bar{\delta}_H = 1$ in (16) and solve it for the pressure distribution. In general, the pressure at the contact line will diverge to positive or negative infinity. Then, the Hertz-like solution is obtained by varying the normalized indentation depth $\bar{\delta}_H$ until the pressure at the contact line is bounded. The corresponding Hertz-like load $\bar{P}_H$ is determined by numerically integrating the pressure distribution using (15b).

Using the fact that the *normalized* Hertz-like load and displacement are functions only of $\beta$, i.e.,

$$\delta_H = (a^2/R)\bar{\delta}_H(\beta); \quad P_H = (16\mu a^3/3R)\bar{P}_H(\beta), \tag{17a,b}$$

the instantaneous compliance can be found using (1) resulting in

$$C = \frac{1}{8\mu a}\phi(\beta), \quad \phi(\beta) \equiv \frac{\bar{\delta}_H - \dfrac{\beta}{2}\dfrac{d\bar{\delta}_H}{d\beta}}{\bar{P}_H - \dfrac{\beta}{3}\dfrac{d\bar{P}_H}{d\beta}}. \tag{18a,b}$$

Here $\phi$ is the normalized compliance. Recall that $1/(8\mu a)$ is the instantaneous compliance of the classical Hertz theory (with no surface tension) so the normalized compliance specifies the deviation due to substrate-air surface tension.

### 3.2 Asymptotic results

Of particular interest is the transition from the elasticity dominated limit to the surface tension dominated limit. It is therefore appropriate to study the behavior of the normalized Hertz-like load and displacement in these limits. The behavior of these functions in the elastic dominated limit is a direct consequence of our choice of normalization, i.e.,

$$\bar{P}_H(\beta \to 0) = \bar{\delta}_H(\beta \to 0) = 1. \tag{19}$$

Equation (18b) and (19) imply that the normalized compliance $\phi(\beta \to 0) = 1$, as expected. It is also possible to show that the integral equation (16), in the limit of $\beta \to 0$, reduces to the integral equation governing the classical Hertz theory without surface tension.

The surface tension dominated or non-wetting liquid like (no adhesion) limit requires more analysis. Using properties of special functions and after some calculations, the integral equation (15a) is reduced to (see supplementary material for details):

$$\frac{4}{\pi\beta}\int_0^{\bar{r}} \bar{t}\bar{p}(\bar{t},\beta)\ln(\bar{r}/\bar{t})d\bar{t} = \bar{r}^2/2, \quad \beta \to \infty. \tag{20}$$

Since the right hand side of (20) does not go to zero as $\beta \to \infty$, we conclude that there exists a function $\chi$ which depends *only* on $\bar{t}$, such that

$$\bar{p}(\bar{t},\beta \to \infty) = \chi(\bar{t})\beta. \tag{21}$$

Furthermore, it is easy to verify that $\chi(\bar{t}) = \pi/2$ satisfies the integral equation (20) exactly. Substituting (21) into (15b), we have

$$\bar{P}_H(\beta \to \infty) = 3\beta\int_0^1 \bar{t}\chi(\bar{t})d\bar{t} \equiv \frac{3\pi}{4}\beta \Leftrightarrow P_H(\beta \to \infty) = 2\pi\sigma a^2/R. \tag{22}$$

Reverting to dimensional quantity, (21) implies that the pressure distribution in the contact zone is

$$p(t,\beta \to \infty) = 2\sigma/R, \tag{23}$$

which is recognizable as the (uniform) Laplace pressure required to balance the force of surface tension. This result is consistent with the fact that in the surface tension dominated limit and in the absence of adhesion, it is the surface tension that fully supports the applied force. The behavior of the Hertz-like displacement can be found by substituting (21) into (15c), it is:

$$\bar{\delta}_H(\beta \to \infty) \to \ln\beta \Rightarrow \delta_H = \frac{P_H}{2\pi\sigma}\ln\left(\frac{\sigma}{2\mu a}\right). \tag{24}$$

The normalized compliance $\phi(\beta)$ is computed using (18b), (22) and (24) and is given by

$$\phi(\beta \to \infty) \to \frac{2}{\pi\beta}\ln\beta - \frac{1}{\pi\beta} \Rightarrow \frac{1}{2\pi a}\frac{dC}{da} = -\frac{1}{4\pi^2\sigma a^2}. \qquad (25)$$

The energy release rate in the surface tension dominated limit is obtained using (2a), (25) and (22) and is:

$$G = \frac{(P - 2\pi\sigma a^2/R)^2}{8\pi^2\sigma a^2}. \qquad (26)$$

The relation between applied load and contact radius can be obtained using the energy balance equation $G = W_{ad}$, which results in (8).

### 3.3 General relation between load and contact radius for the adhesive problem

The relation between load and contact radius is given by setting the energy release rate given by (2a) equal to the interfacial work of adhesion. It is convenient to introduce a new normalization for the applied load that is independent of the unknown contact radius $a$:

$$\hat{P} = \frac{P}{\frac{3}{2}\pi R W_{ad}}, \quad \hat{a} = \frac{a}{\frac{1}{2}\left(\frac{9\pi W_{ad}}{4\mu}\right)^{1/3} R^{2/3}}. \qquad (27a,b)$$

Here, the indentation load $P$ and contact radius $a$ are normalized by the magnitude of pull-off load and pull-off contact radius in the JKR limit (see 10(a, b)). With this normalization, (2a) and the energy balance equation $G = W_{ad}$ become (see supplementary material for details):

$$\hat{P} = \overline{P_H}(\beta)\hat{a}^3 - 2\sqrt{\frac{\hat{a}^3}{\Lambda(\beta)}}, \qquad (28a)$$

where,

$$\Lambda(\beta) \equiv -16\pi\mu a^3 \frac{dC}{dA} = \phi(\beta) + \beta\frac{d\phi(\beta)}{d\beta}. \qquad (28b)$$

where $A = \pi a^2$ is the contact area. Although $\beta$ is convenient to use from a theoretical standpoint, it depends on the contact radius, which is usually an unknown. Hence we introduce a new dimensionless parameter

$$\omega = \frac{\sigma}{(\mu R)^{2/3}\left(\frac{9\pi}{4}W_{ad}\right)^{1/3}} = \beta\hat{a}. \qquad (29a)$$

Equation (28a) is a key result: it generalizes the JKR theory to include surface tension. The modified theory is simple, in that two scalar factors $\overline{P_H}(\beta)$ and $\Lambda(\beta)$ that depend only on $\beta$ need to be appended to the JKR theory to account for surface tension. Since $\beta = \omega/\hat{a}$, it implies that the relation between the normalized contact radius and the normalized applied load depends on a *single* dimensionless parameter $\omega$. This dimensionless parameter was also found in previous studies using molecular dynamics for the case of zero indentation load[7]. The same parameter also governs the problem that a soft particle is in contact with a rigid substrate[9,11,21,22].

### 3.4 Numerical Results: $\bar{P}_H(\beta), \bar{\delta}_H(\beta), \phi(\beta)$ and $\Lambda(\beta)$

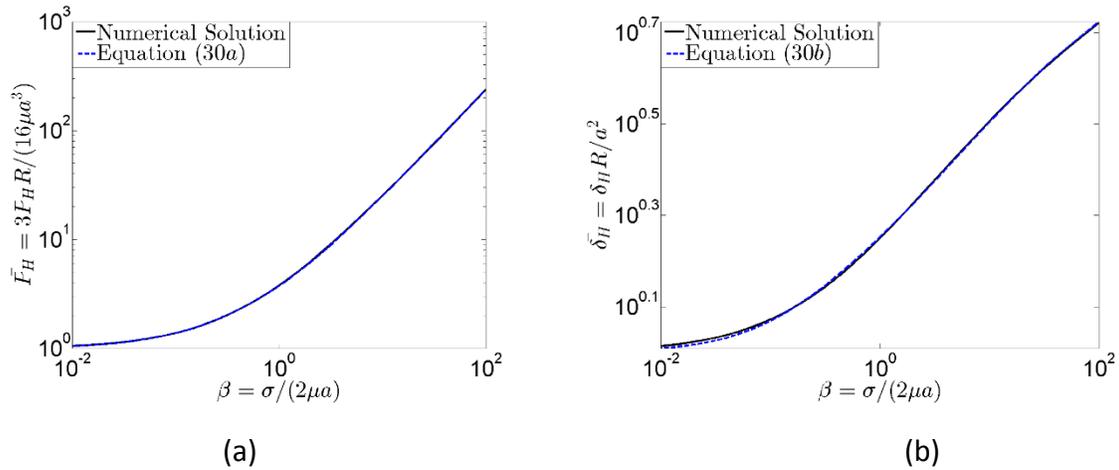

(a)                                  (b)

Fig. 4 The vertical axis is the ratio of the Hertz like load/displacement with surface tension to the classical Hertz load/displacement without surface tension. The horizontal axis is the normalized surface tension $\beta$. These figures quantify the deviation from classical Hertz theory due to air-substrate surface tension. (a) Normalized Hertz-like load $\bar{P}_H$ against $\beta$ (b) Normalized Hertz-like displacement $\bar{\delta}_H$ against $\beta$

Fig. 4(a,b) plot the normalized Hertz-like load $\bar{P}_H$ and normalized Hertz-like displacement $\bar{\delta}_H$ against $\beta$. As shown in Fig. 4(a,b), these numerical results are very well numerically approximated by

$$\bar{P}_H = 1 + \frac{3\pi\beta}{4}\left[\frac{\beta^2 + 0.6016\beta + 0.0171}{\beta^2 + 0.3705\beta + 0.0063}\right], \tag{30a}$$

$$\bar{\delta}_H = 1 + \frac{1}{3}\ln(1 + 6.582\beta + 2.759\beta^2 + 0.3782\beta^3). \tag{30b}$$

It should be noted that (30a,b) give the correct asymptotic behaviors for small and large $\beta$. Equations (30a,b), together with (18b) and (28b) allow us to compute the scalar factor $\Lambda(\beta)$. Since the calculation involves second derivatives of $\bar{P}_H$ and $\bar{\delta}_H$, a local fit method is used to determine $\phi(\beta)$ and $\Lambda(\beta)$. The results are shown in Fig. 5. Also, for the sake of completeness, we plotted the normalized pressure distribution of the Hertz –like problem for different values of $\beta$. These results are given in the supplementary materials.

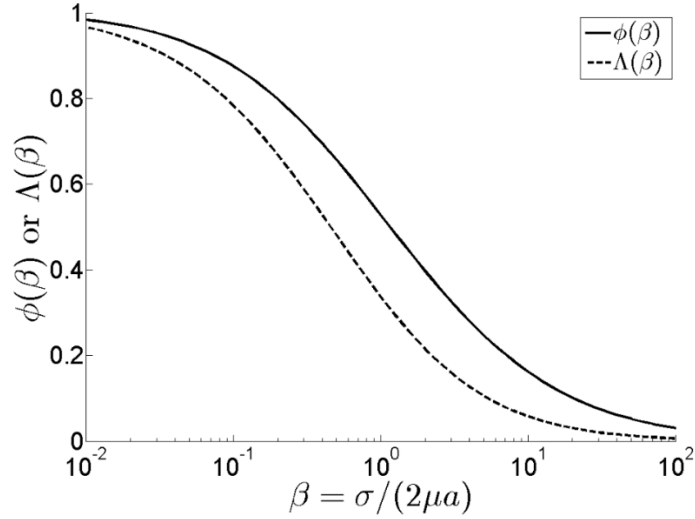

Fig. 5 Dimensionless compliance $\phi(\beta)$ and $\Lambda(\beta)$ versus $\beta$

$\phi(\beta)$ and $\Lambda(\beta)$ can be numerically approximated by:

$$\phi(\beta) = \frac{0.6425\beta^2 + 4.284\beta + 1}{0.2039\beta^3 + 3.994\beta^2 + 6.011\beta + 1} \tag{31a}$$

$$\Lambda(\beta) = \frac{2\beta^2 + 7.4\beta + 1}{\pi\beta^3 + 15.78\beta^2 + 10.92\beta + 1} \tag{31b}$$

Again, these approximations give correct asymptotic behaviors for both small and large $\beta$.

### 3.5 Pull-off and zero load cases

Of particular interest are the contact radius at zero load $(a_0)$ and the pull-off force $(P_{off})$, which can be readily determined from (28a). The contact radius at zero load satisfies the implicit equation

$$\hat{a}_0 = \left( \frac{4}{\Lambda(\omega/\hat{a}_0)\left[\bar{P}_H(\omega/\hat{a}_0)\right]^2} \right)^{1/3}. \tag{32}$$

The normalized pull-off force $\hat{P}_{off}$ and the contact radius at pull-off $\hat{a}_{off}$ are determined from (28a) using the condition $d\hat{P}/d\hat{a} = 0$ and they are functions of only one dimensionless parameter $\omega$. Since $P$ and $a$ are normalized by the magnitude of pull-off load and pull-off contact radius in the JKR limit, the results for $\hat{P}_{off}$ and $\hat{a}_{off}$ reflect the deviation of pull-off behavior from the classical JKR theory due to surface tension. These results are shown in Fig. 6.

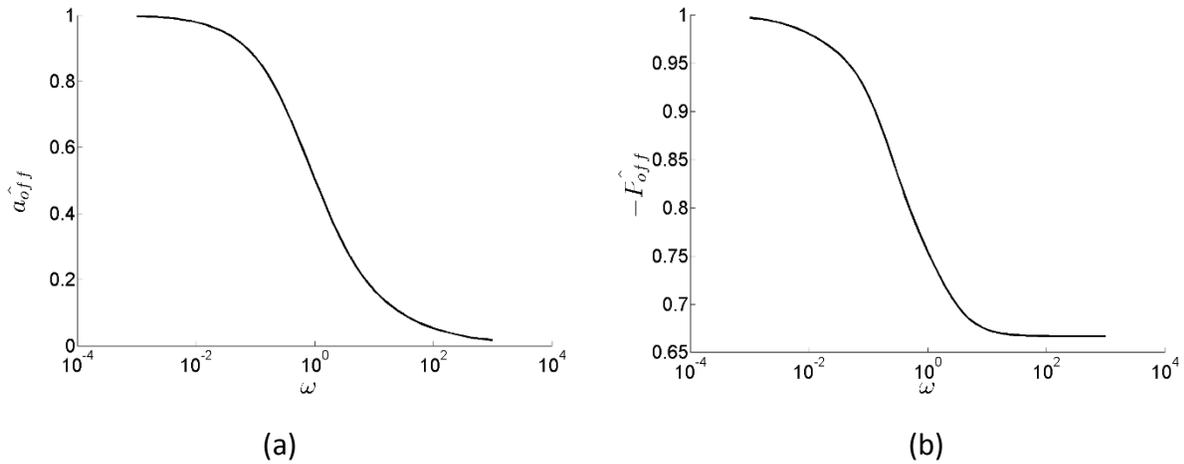

(a)          (b)

Fig. 6 (a) Normalized contact radius at pull-off versus $\omega$. (b) Normalized pull-off force versus $\omega$.

### 3.6 Contact radius versus applied load

The contact radius versus load relation for various $\omega$ is shown in Fig. 7. Note that the small contact requires $W_{ad}/(\mu R)$ to be much less than $\max\{1, \sigma/(\mu R)\}$. As expected, increasing $\omega$ (e.g. increasing surface tension at fixed modulus, sphere radius and work of adhesion) increases the substrate's resistance to deformation, resulting in much smaller contact at the same indentation load.

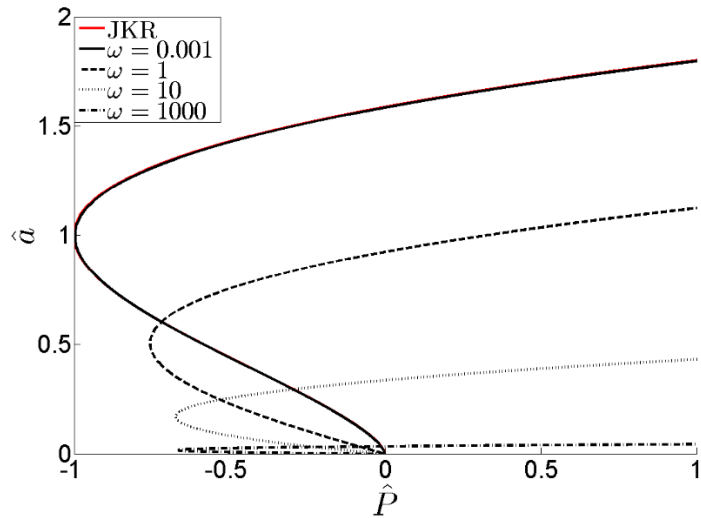

Fig. 7 Normalized contact radius versus normalized load for different $\omega$

### 3.7 Indentation depth versus contact radius

In a displacement controlled test, the relation between indentation depth and contact radius is obtained using (2b), (18a,b), and (28b):

$$\hat{\delta} = 3\hat{a}^2 \bar{\delta}_H(\beta) - 4\phi(\beta)\sqrt{\frac{\hat{a}}{\Lambda(\beta)}} ,\qquad(33)$$

where $\hat{\delta}$ is the indentation depth normalized by the magnitude of pull-off displacement in the JKR limit and $\bar{\delta}_H$ is approximated by (30b). Equation (33) reduces to the JKR theory (3c) in the limit of $\beta \to 0$ since $\bar{\delta}_H(\beta), \phi(\beta), \Lambda(\beta)$ approach 1 in this limit. The contact radius versus indentation depth for different values of $\omega$ is shown in Fig. 8.

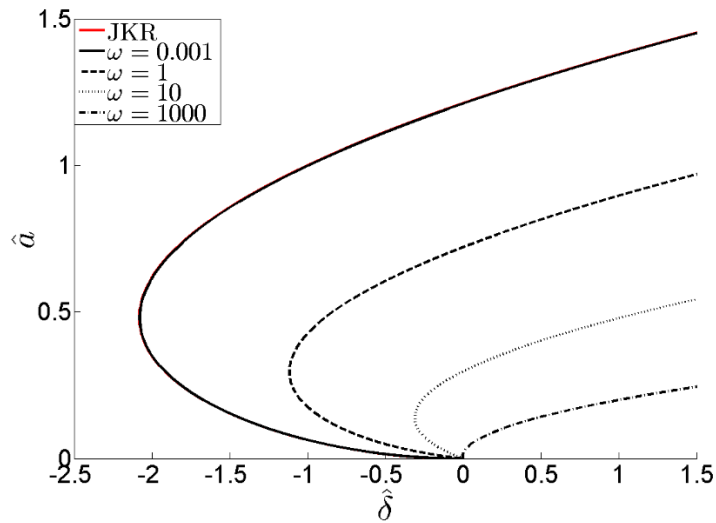

Fig. 8 Normalized contact radius versus normalized indentation depth for 4 different $\omega$. For $\omega = 0.001$ the JKR solution and the prediction given by (33) lie on top of each other.

Of great interest to experimentalists is the relation between applied load and indentation depth, which can be obtained by combining the results from Fig. 7 and Fig. 8 and is shown in Fig. 9.

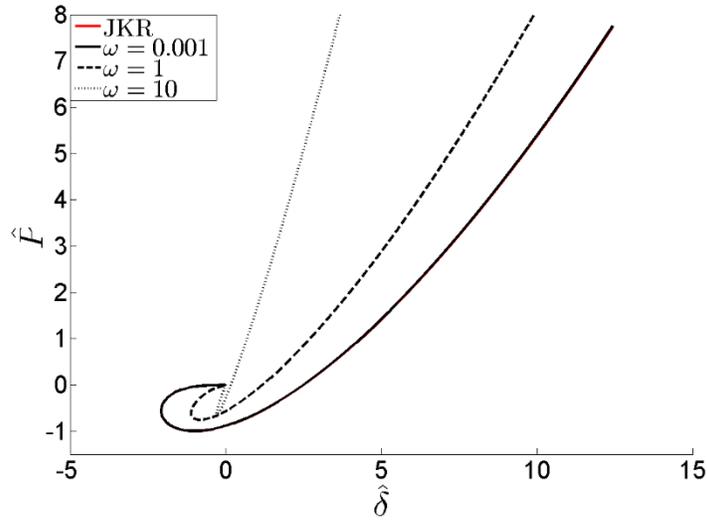

Fig. 9 Normalized load versus normalized indentation depth for different $\omega$. Note that JKR solution and the solution for $\omega = 0.001$ lie on top of each other.

## 4. Comparison with experiments

Style et al. [5] have recently measured the contact radius of small glass spheres adhering to the surface of silicone substrates under zero external load (the sphere are sufficiently small so that adhesive forces dominate gravity). Different contact radii are obtained by varying the radius of the sphere (from 3 to 30 $\mu m$) and the substrate's shear modulus $\mu$ (from 1 to 167kPa). Their results (symbols) are shown in Fig. 10. Xu et al. [6] previously analyzed the special case of the problem considered in the present work when force is zero, using a large-strain finite element (FEM) method. Their results were validated by fitting experimental data of Style et al. [5], with work of adhesion and surface tension extracted as two fitting parameters. While the analytical model developed in the present work extends the previous FEM analysis by allowing for applied indentation force, it is restricted by the assumptions that strains and displacements remain small. In Fig. 10, using the same parameters as found by Xu et al. [6], we compare the results of the small-strain theory developed in the present work with experiments of Style et al. [5]. The solid lines represent the full theory; the dotted lines represent the theory when air-substrate surface tension is neglected (classical JKR solution). Note that for the three stiffer materials, the present small-strain theory fits the data very well, even though the deformation is not that small. There is a systematic difference between results that do or do not include the effect of surface tension. For the most-compliant substrate, the small-strain theory significantly overestimates the contact radius. The breakdown of small strain theory in this regime is not surprising, since the small strain theory is expected to break down for this case except close to the non-wetting limit, where $W_{ad}/\sigma \ll 1$. To confirm that this discrepancy is due only to the very large deformation in the experiment, we modified our derivation of (8) without making the small angle approximation and obtained the following relation between load and contact radius in the partial-wetting liquid-like limit:

$$P = \frac{2\pi\sigma a^2}{R} - 2\pi a\sigma\left[\frac{a\,W_{ad}}{R\sigma} + \sqrt{1-(a/R)^2}\sqrt{1-(1-W_{ad}/\sigma)^2}\right]. \tag{34}$$

Note that, if the work of adhesion is indeed small in comparison with the surface tension and $a/R \ll 1$, then (33) reduces to the small strain theory solution given by (8). For the special case of $P = 0$, (33) reduces to a result in [6]:

$$a_0 = R\sqrt{1-(1-W_{ad}/\sigma)^2}. \tag{35}$$

This equation is used to fit the data in Fig. 10 for $\mu = 1kPa$. The almost perfect fit between model and experiments supports our claim that for very soft and adhesive samples the small deformation theory breaks down.

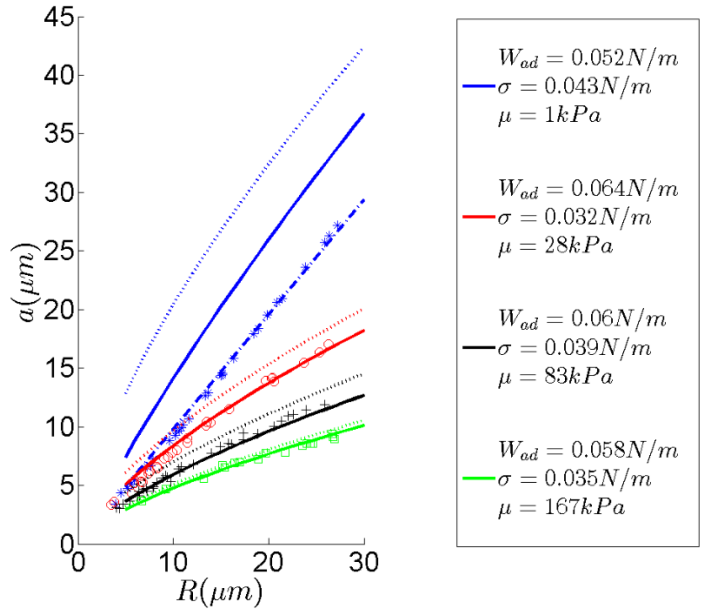

Fig. 10 Comparison with experimental data of Style *et al.* [5]. Symbols are experimental data. Solid lines are predictions based on small deformation theory with air-substrate surface tension. Dotted lines are based on small deformation theory without surface tension. Prediction of large strain solution in partial wetting liquid-like limit (34) is shown as a dashed-dotted line. The parameters used for these fits are based on the paper by Xu *et al.* [6].

## 5. Summary and Discussion

Equations (28a) and (33) extend the JKR theory to include the effect of surface tension. The extensions involve three dimensionless scalar factors, $\bar{P}_H, \bar{\delta}_H, \Lambda$, which depend on the elasto-capillary number $\beta = \sigma/(2\mu a)$ only. The dimensionless parameter

$$\omega = \frac{\sigma}{(\mu R)^{2/3}\left(\frac{9\pi}{4}W_{ad}\right)^{1/3}}$$ completely determines the indentation load-contact radius relations and the pull-off behavior. Except for a numerical constant, $\omega$ is the ratio of the elasto-capillary length $\sigma/\mu$ to a characteristic length $R^{2/3}(W_{ad}/\mu)^{1/3}$ where adhesion acts (the pull-off contact radius).

Our analysis is strictly valid for a rigid sphere in no-slip contact with an incompressible isotropic elastic substrate. The incompressibility of the substrate is needed to ensure the vanishing of shear stress in the contact region. However, this incompressibility requirement can be relaxed since the interfacial shear stress due to a compressible substrate is typically small in comparison with the contact pressure, as noted by Johnson [1]. Therefore, our results are also approximately valid for compressible substrates provided that the shear modulus $\mu$ in our equations is replaced by $\mu[2(1-v)]^{-1}$.

Our analysis uses small deformation theory which requires the contact radius to be much less than the radius of the sphere. However, for soft materials the deformation can be very large. Fortunately, the works of Xu *et al.* [6] and Lin *et al.* [23] showed that the results of small deformation theory are surprisingly robust for this class of contact problems. This claim is also supported by our result in Fig. 10 which shows that small deformation theory is surprisingly accurate as long as the elasto-capillary number is not so large that one approaches the partial wetting liquid-like limit. The smallest elasto-capillary number $\alpha = \sigma/(2\mu R)$ corresponding to the case of $\mu = 1kPa$ in Fig. 10 is about 0.7. Note that small strain theory is valid even though $\alpha \geq 1$ as long as $W_{ad}/\sigma \ll 1$. If the condition $W_{ad}/\sigma \ll 1$ is violated and $\alpha \geq 1$, we suggest using (34) to approximate the relation between the indentation load and the contact radius.

Our calculation is strictly valid for no slip contact. For substrates such as hydrogels, the frictionless boundary condition could be more appropriate. Recall that the classical JKR theory is strictly valid for frictionless and no slip contact, as long as the substrate is incompressible. However, as pointed out in section 2, without the no-slip condition, one would have to introduce new physics into the model – that is, the tension of the substrate surface will change since it is in contact with the rigid surface. The problem is that the interfacial tension between a rigid surface and the surface of a compliant solid is not well defined, as a rigid surface can support any tension. From the mechanics view point, there is no consistent way to assign the tension acting on the substrate surface after contact. One way to deal with this difficulty is to assign the same tension to the substrate surface after contact, which is essentially the approach of Long *et al*. [19]. In this case the results of the present article apply without modification. However, the effect of this approximation could result in a work of adhesion that is different from that given by the Young-Dupre equation. This ad-hoc assumption is supported, at least partially by the fact that our theory is able to match the data of Style *et al*. [5].

We have not found any experiments that specifically measures the effect of surface tension on the relation between indentation and force. Therefore, our comparison is based on the special experimental case where the applied force is zero. It will be of great interest to test this theory against experiments involving finite indentation force.

**Ethics Statement:** This work do not involved human subjects or animals.

**Data Accessibility:** All data are generated by numerical solution of integral equation which is described in main text and in supplementary materials.

**Competing interest:** We have no competing interest.

**Authors' contributions:** The problem is conceived during a discussion with E. Raphael, T. Salez and A. Jagota. C.Y. Hui formulated the problem and drafted the initial manuscript, with valuable physical insights from E. Raphael, T. Salez and A. Jagota. All numerical calculations were done by T. Liu and T. Salez. All authors gave final approval for publications.

**Acknowledgements:** The authors would like to thank Michael Benzaquen for interesting discussions.

**Funding:** C.Y. Hui, A. Jagota and T. Liu acknowledge support from the U.S. Department of Energy, Office of Basic Energy Science, Division of Material Sciences and Engineering under Award (DE-FG02-07ER46463).

<u>Supplementary Material</u>

**Numerical Implementation**

We divide the interval $[0,1]$ into $n$ equal subintervals $[t_i, t_{i+1}]$, $i=1,....,n+1$ with equal length $\Delta = 1/n$, so $t_i = (i-1)\Delta$, $i=1,....,n+1$. Let the midpoints of these intervals to be denoted by

$$\frac{t_j + t_{j+1}}{2} = r_j = \frac{2j-1}{2}\Delta, \quad j=1,...,n \tag{S1}$$

We discretize (16) into a $n \times n$ linear system of equations,

$$\frac{2}{\pi\beta}\int_0^1 \overline{tp}(\overline{t},\beta)\int_0^\pi \psi(\overline{l}/\beta)d\theta d\overline{t} = \sum_{i=1}^n M_{ji}\overline{p}_i = \overline{\delta}_H - \frac{\overline{r}_j^2}{2} \quad j=1,...,n \tag{S2}$$

where $\overline{p}_i = \overline{p}(\overline{t} = \overline{r}_i, \beta)$ and $M_{ji}$ is the $n \times n$ matrix defined by

$$M_{ji} = \frac{2}{\beta\pi}\int_{t_i}^{t_{i+1}} \overline{t}k(\overline{t},\overline{r}_j,\beta)d\overline{t}, \quad k(\overline{t},\overline{r},\beta) = \int_0^\pi \psi\left(\sqrt{\overline{t}^2+\overline{r}^2-2\overline{rt}\cos\theta}/\beta\right)d\theta \tag{S3}$$

We find n = 20 is enough to give us good results. The coefficients $M_{ji}$ are obtained using standard quadrature rules. For any given $\beta$, the initial value of $\overline{\delta}_H$ is taken to be 1, and it is updated using the following shooting scheme:

Step 1: Solve (S2) using $\overline{\delta}_H = \overline{\delta}_1 = 1$.

Step2: Solve (S2) using $\overline{\delta}_H = \overline{\delta}_2 = 2\overline{\delta}_1$.

Step3: Examine the divergence of the pressure field near the contact line in step 1 and step 2. If pressure diverges to positive infinity at $\overline{\delta}_1$ and to negative infinity at $\overline{\delta}_2$, then the correct $\overline{\delta}_H$ must lie in between $\overline{\delta}_1$ and $\overline{\delta}_2$, then go to step 4. If both $\overline{\delta}_1$ and $\overline{\delta}_2$ lead to positive infinite pressure, then take $\overline{\delta}_1 = \overline{\delta}_2$ and go back to step 2.

Step4: Let $\overline{\delta}_3 = \frac{|d\overline{p}_2|\overline{\delta}_1 + |d\overline{p}_1|\overline{\delta}_2}{|d\overline{p}_1| + |d\overline{p}_2|}$, $d\overline{p}_1 = \overline{p}_n(\overline{\delta}_1) - \overline{p}_{n-1}(\overline{\delta}_1)$, $d\overline{p}_2 = \overline{p}_n(\overline{\delta}_2) - \overline{p}_{n-1}(\overline{\delta}_2)$. Solve (S2) using $\overline{\delta}_H = \overline{\delta}_3$. Check whether the pressure converges at the edge (r=a). If not, take $\overline{\delta}_1 = \overline{\delta}_3$ if it diverges to positive infinity or take $\overline{\delta}_2 = \overline{\delta}_3$ if it diverges to negative infinity and then repeat step 4. During this loop, always check whether pressure converges for $\overline{\delta}_1$ and $\overline{\delta}_2$. Whenever it converges we get the correct answer for both $\overline{\delta}_H$ and $\overline{P}_H$.

For each $\beta$, the numerical solution gives us the correct normalized displacement $\overline{\delta_H}(\beta)$, which should be a decreasing function of $\beta$. The total normalized Hertz-like force is finally computed using (15b).

The pressure distributions for different $\beta$ are shown as below:

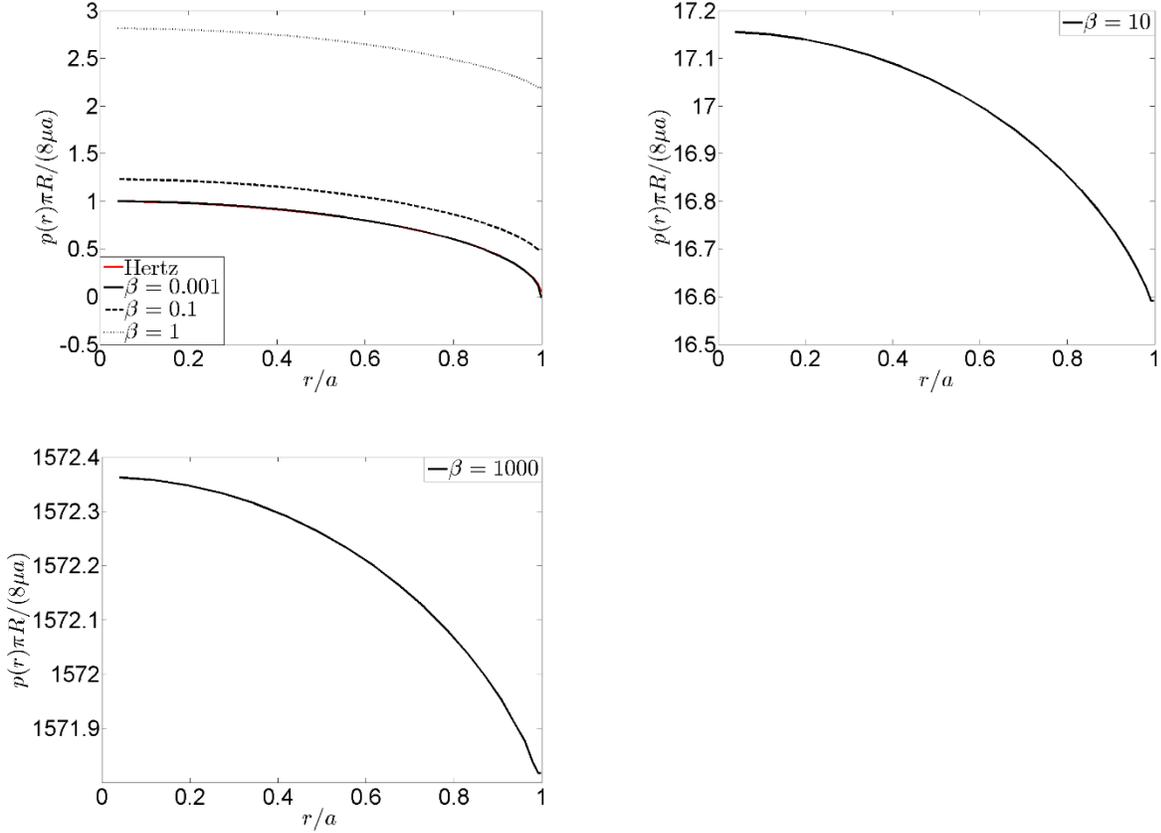

### Derivation of Equations (20) and (28)

According to the properties of Bessel function of the second kind and Struve function,

$$\psi(x \to 0) \approx -\frac{2}{\pi}[\ln(\frac{x}{2}) + \gamma],\qquad\text{(S4)}$$

where $\gamma$ is the Euler constant. Using (S4),

$$\psi(\bar{t}/\beta) - \frac{1}{\pi}\int_0^\pi \psi(\bar{l}/\beta)d\theta \approx -\frac{2}{\pi}[\ln(\bar{t})] - \frac{1}{\pi}\int_0^\pi \ln(\sqrt{\bar{r}^2 + \bar{t}^2 - 2\bar{r}\bar{t}\cos\theta})d\theta]$$

$$= \begin{cases} 0, & \bar{r} < \bar{t} \\ \frac{2}{\pi}\ln(\frac{\bar{r}}{\bar{t}}), & \bar{r} > \bar{t} \end{cases} \qquad\text{(S5)}$$

Equation (20) is obtained by substituting (S5) into (15a).

To derive (28), we equate the energy release rate from (2a) to the work of adhesion, resulting in

$$-\frac{1}{2}\frac{dC}{dA}(P-P_H)^2 = W_{ad}, \qquad (S6)$$

where $A = \pi a^2$ is the contact area. Using the normalization of (14) and (27), (S6) becomes:

$$-\frac{1}{2}\frac{dC}{dA}(P-P_H)^2 = -\frac{1}{2}\frac{\Lambda(\beta)}{-16\pi\mu a^3}(\frac{16\mu R^2}{3})^2(\hat{P}-\overline{a}^3\overline{P_H})^2 = \frac{16\mu R}{9\pi\overline{a}^3}\frac{1}{2}\Lambda(\beta)(\hat{P}-\overline{a}^3\overline{P_H})^2 = W_{ad}. \qquad (S7)$$

After rearrangement, (S7) becomes (28a).